# Magnetization jumps and relaxation effect in doped CeFe$_2$

Arabinda Haldar[1], K. G. Suresh[1,3] and A. K. Nigam[2]
[1]Magnetic Materials Laboratory, Department of Physics
Indian Institute of Technology Bombay, Mumbai- 400076, India
[2]DCMPS, Tata Institute of Fundamental Research,
Homi Bhabha Road, Mumbai- 400005, India

E-mail:suresh@phy.iitb.ac.in

**Abstract**. For the first time, we find that the dynamic antiferromagnetic phase present in CeFe$_2$ gets stabilized with Ga and Si substitutions. We find that phenomena such as strain-induced first order jumps in the magnetization curves, asymmetry between the *M-H* curves during the increasing and decreasing field cycles, the envelope curve being inside the virgin curve, occur in these compounds. Temperature and time dependences of magnetization show that the compounds possess glassy behavior at low temperatures. Multi-step magnetization behavior, unusual relaxation effect, thermal and magnetic history dependence, which are signatures of the martensitic scenario due to the strong magneto-structural coupling, are found to be present in this system. We also show that one can induce the magnetization steps with the help of appropriate measurement protocol. Detailed magnetization relaxation studies have been carried out to understand the dynamics of magnetic phase transition.

## 1. Introduction

Doped CeFe$_2$ is a well known phase separated compound among the intermetallics [1]-[4]. Both structural and magnetic phase transitions and a strong coupling between these two has made this system a prototype to address distinct features of basic magnetism, kinetics and first order phase transition [5]-[7]. CeFe$_2$ shows simple ferromagnetic behavior with a Curie temperature (T$_C$) of 230K and it is shown that antiferromagnetic spin correlation exists at low temperatures [8]. However, application of external hydrostatic pressure or substitution of Fe by some selected elements stabilizes the antiferromagnetic ground state in this compound [9]. Neutron diffraction studies on the doped CeFe$_2$ have shown that the compound undergoes a structural distortion while going from high temperature ferromagnetic phase (cubic) to low temperature antiferromagnetic phase (rhombohedral) [1]. Ultra sharp steps in magnetization isotherms have been attributed to the martensitic behavior in these compounds [7]. It is well known that certain phase separated manganites undergo structural transition across the magnetic transition [10]-[12]. There seems to be many similarities between doped CeFe$_2$ and these manganites. Taking into account these similarities, martensite type behavior is invoked to address the anomalous features shown by doped CeFe$_2$ compounds.

---

[3] Author to whom any correspondence should be addressed.

In this report, we show that substitution of Fe by Ga and Si can stabilize the low temperature antiferromagnetic state when the doping concentration is above a critical value. As representative examples, we focus on $Ce(Fe_{0.975}Ga_{0.025})_2$ and $Ce(Fe_{0.95}Si_{0.05})_2$, in which the AFM state is quite stable at low temperatures. Martensitic scenario and kinetics of the system has been discussed with the help of detailed magnetization relaxation measurements. Universality of these features and that of magnetic oxides has been discussed.

## 2. Experimental Details

Details of preparation of polycrystalline samples of $CeFe_2$, $Ce(Fe_{1-x}Ga_x)_2$, $Ce(Fe_{1-x}Si_x)_2$ [$x$=0, 0.01. 0.025 and 0.05] have been reported elsewhere [1,7]. The structural analysis of the samples was performed by collecting the room temperature powder x-ray diffractograms (XRD) using Cu-K$_\alpha$ radiation. The refinement of the diffractograms was done by the Rietveld analysis using *Fullprof* suite program. The lattice parameters were calculated from the refinement. The DC magnetization measurements, in the temperature range of 1.8- 300 K and in fields up to 90 kOe have been performed in a Vibrating Sample Magnetometer attached to a Physical Property Measurement System (PPMS, Quantum Design Model 6500) and/or a SQUID magnetometer. Some measurements were done using Oxford Maglab VSM. Magnetization has been measured in zero field cooled (ZFC), field cooled cooling (FCC) and field cooled warming (FCW) modes.

## 3. Results and Discussions

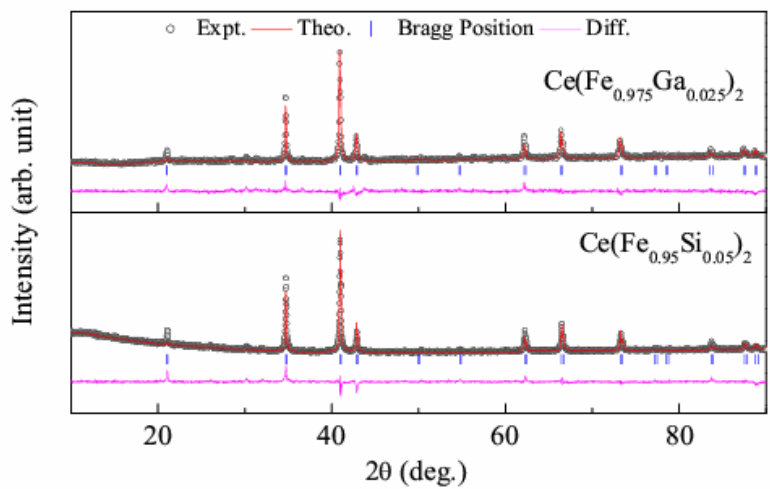

**Figure 1**. Powder x-ray diffraction patterns, along with the Rietveld refinement of $Ce(Fe_{0.975}Ga_{0.025})_2$ and $Ce(Fe_{0.95}Si_{0.05})_2$ compounds. The plots at the bottom show the difference between the theoretical and the experimental data.

X-ray diffraction patterns of $Ce(Fe_{0.975}Ga_{0.025})_2$ and $Ce(Fe_{0.95}Si_{0.05})_2$ are shown in figure 1 along with the Rietveld refinement. It is clear that the compounds are single phase and both possess $MgCu_2$ type cubic structure (Space group: $Fd\overline{3}m$ ).

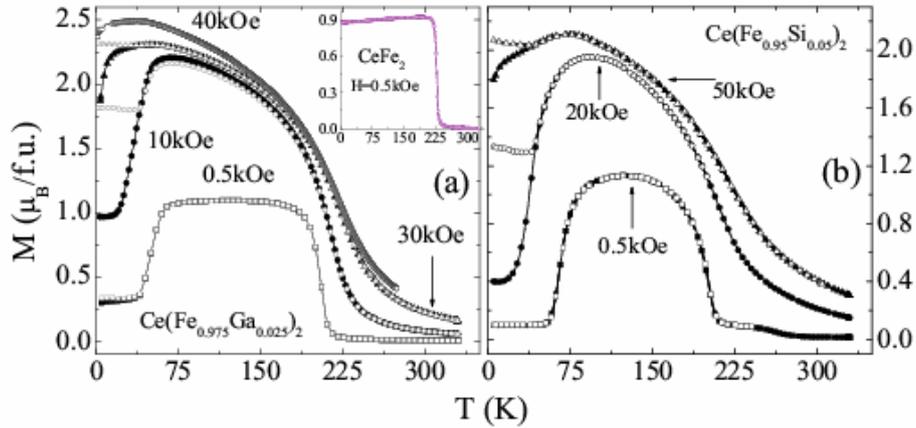

**Figure 2**. Temperature dependence of the magnetization of (a) Ce(Fe$_{0.975}$Ga$_{0.025}$)$_2$, (b) Ce(Fe$_{0.95}$Si$_{0.05}$)$_2$ in various fields. Data has been taken during warming the sample in both ZFC (closed symbols) and FCW (open symbols) modes. Inset shows the M vs T plot for CeFe$_2$.

Temperature variation of magnetization is shown in figure 2 for both the compounds. The inset of figure 2a shows the M-T data of CeFe$_2$. With substitution of Ga and Si the antiferromagnetic phase gets stabilized at low temperatures. A similar observation was also reported with Co, Al, Ru, Ir, Os and Re doping [1]-[4]. The difference between ZFC and FCW data increases with increasing field and at high enough fields, these two modes of data merge. It is to be noted that while Ga concentration of 0.025 stabilizes the AFM state very well, it needs a higher concentration of 0.05 in the case of Si substitution. Comparing different concentrations of Si and Ga (all not shown in this paper) above the critical concentration, it is found that Si doped compounds possesses stronger antiferromagnetic coupling at low temperatures, compared to the Ga doped compounds.

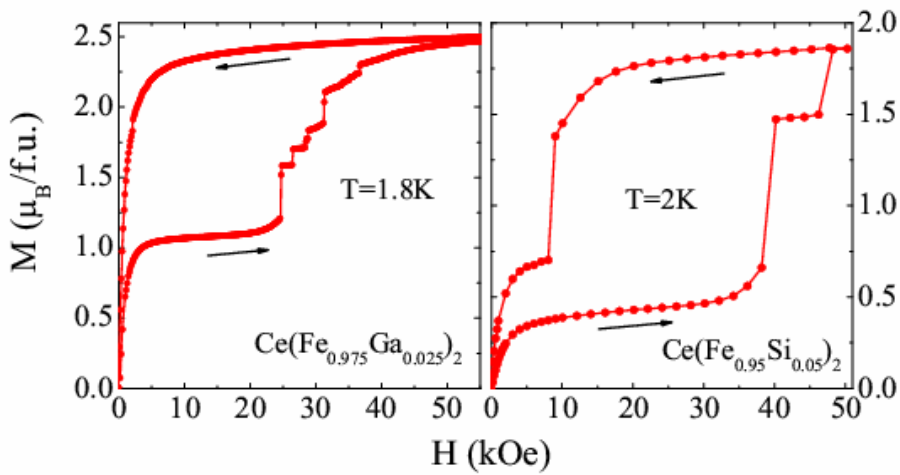

**Figure 3**. Magnetization isotherms for Ce(Fe$_{0.975}$Ga$_{0.025}$)$_2$ at 1.8K and Ce(Fe$_{0.95}$Si$_{0.05}$)$_2$ at 2K.

Magnetization isotherms at low temperature are shown for both the compounds in figure 3. It is interesting to note here that across the metamagnetic transition the moments reorient themselves along the direction of field, accompanied by a number of sudden jumps. The field decreasing path does not follow the field increasing path in the magnetization, for both the compounds. The difference between the two branches was earlier attributed to the supercooling of the ferromagnetic phase [5]. To account for these jumps, we would recall the structural transition across the magnetic phase transition from AFM to FM states (with the help of the applied field), which was reported earlier in other doped $CeFe_2$ compounds [1]. Structural mismatch across the magnetic phase transition produces martensitic strains in the compound and during field induced transition the strain gets released along with the spin flip. This relief of strain energy along with magnetic phase transition causes the compound to transform from AFM phase to FM phase in a burst-like manner, resulting in the jumps [7], [10]-[12]. It may also be noted from figure 3 that the critical field for metamagnetic transition is more in the case of Si doped compound, as compared to that in the Ga doped compound. This observation is in agreement with the fact that the AFM is better stabilized with Si.

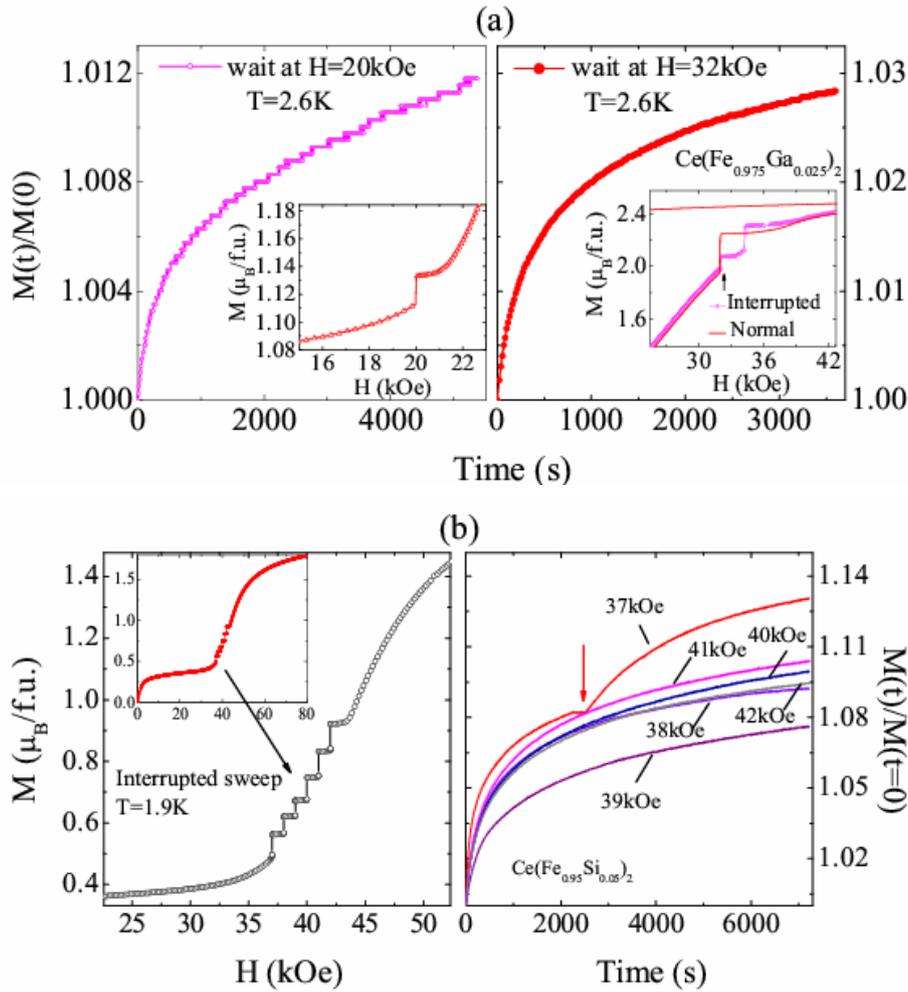

**Figure 4.** (a) Magnetization isotherm at 2.6K for $Ce(Fe_{0.975}Ga_{0.025})_2$ in the interrupted sweep mode. Measurement was held for 1.5 h at 20kOe and 1 h at 32kOe. (b) mageization isotherm at 1.9K for $Ce(Fe_{0.95}Si_{0.05})_2$ compound. Measurement was done for holding time of 2 h at fields of 37, 38, 39, 40, 41 and 42 kOe. Right panels of a and b show the growth of magnetization with time.

As the magnetic relaxation/growth in these compounds is unconventional, we have studied the time dependence or spontaneous transition across metamagnetic region in detail, as shown in figure 4. For time dependent measurement we have used the following measurement protocol as reported earlier

[13, 14]. The compound has been zero field cooled to the measurement temperature. The field was ramped to the holding field in sweep mode and the measurement was carried out for reasonable duration. Then the field was again ramped to the next holding field and so on. This kind of measurement is referred to here as interrupted sweep mode and is shown in figure 4. It is interesting to observe here that one can induce magnetization steps by delaying the measurement at certain fields across the metamagnetic transition. Time induced magnetization steps have been earlier reported in manganites by holding at a certain field after a critical time [13]-[15]. It shows that the step size depends on the measurement procedure. The experimental time scale and the time scale of transformation may not be same and different measurement protocols give rise to variations in the results. This strong dependence on the measurement protocol does point to the features of martensitic like transformation [12,15]. Therefore, the present observations clearly point towards the martensitic like scenario in doped CeFe$_2$ compounds. Right panels of figure 4a and b show the growth of magnetization during the holding time. It is found that, in general, this variation can be fitted to a stretched exponential function, implying that the magnetic state at low temperatures is of glassy nature [13,16].

The above discussed results indicate the martensitic type nature of the magnetostrucrural phase transition in Ga/Si doped CeFe$_2$. Although both the dopings show stabilization of low temperature AFM phase, the strength of AFM coupling is more in the case of Si doping. The critical fields needed for AFM-FM transition is found to be larger in Si doping as compared to that in Ga doping. Large number of steps are found in Ga doped compound and the metamagnetic transition starts at lower field compared to that of Si doped compound. Another difference between these two substitutions is that the incubation time is less in Ga case as compared to that of Si. Stronger magnetostructural coupling in the case of Si is possibly responsible for longer relaxation time for moments to get saturated at a certain field in the case of Si doped compounds.

## 4. Conclusion

To summarize, we have shown that substitution of Fe by Ga and Si can stabilize the fluctuating AFM ground state in CeFe$_2$. Low temperature magnetic isotherm shows sharp jumps across the metamagnetic transition. Detailed time dependent measurements have been performed to understand the kinetics of the system. Structural mismatch and relief of strain energy across the magnetostructural phase transition have been attributed to the origin of step behavior in magnetization. Relaxation across the metamagnetic transition shows a glassy behavior within this compound. The close similarities between these results and that of manganites are explained in terms of more than one type of competing interactions and their coupling. We conclude that the interesting features in these doped compounds are associated with the first order magnetostructural phase transition.